\documentclass{article}
\usepackage{spconf,amsmath,graphicx}

\usepackage{enumitem}
\usepackage{makecell}
\usepackage{amssymb}
\setlist{nosep, leftmargin=14pt}

\usepackage{mwe} 

\title{Self Pre-training with Adaptive Mask Autoencoders for Variable-Contrast 3D Medical Imaging}

\usepackage{tikz}

\newcommand\copyrighttext{%
  \footnotesize \textcopyright 20XX IEEE. Personal use of this material is permitted. Permission from IEEE must be obtained for all other uses, in any current or future media, including reprinting/republishing this material for advertising or promotional purposes, creating new collective works, for resale or redistribution to servers or lists, or reuse of any copyrighted component of this work in other works.}

\newcommand\mycopyrightnotice{%
\begin{tikzpicture}[remember picture,overlay]
\node[anchor=south,yshift=10pt] at (current page.south) {\fbox{\parbox{\dimexpr0.75\textwidth-\fboxsep-\fboxrule\relax}{\copyrighttext}}};
\end{tikzpicture}%
}

\name{
    \parbox{\textwidth}{
        \centering
        Badhan Kumar Das$^{\ddagger \S \star }$ \qquad Gengyan Zhao$^{\dagger \star}$ \qquad Han Liu$^{\dagger }$ \qquad Thomas J. Re$^{\dagger}$ \\ 
        Dorin Comaniciu$^{\dagger}$ \qquad Eli Gibson$^{\dagger}$ \qquad Andreas Maier$^{\S}$ \\
        \thanks{$^{\star}$ Equal Contribution}
    }
}

\address{$^{\ddagger}$ Siemens Healthineers AG \\
    $^{\dagger}$ Siemens Medical Solutions USA, Inc. \\
    $^{\S}$ FAU Erlangen-Nuremberg
    }

\begin{document}

%
\maketitle

\mycopyrightnotice

\begin{abstract}

The Masked Autoencoder (MAE) has recently demonstrated effectiveness in pre-training Vision Transformers (ViT) for analyzing natural images. By reconstructing complete images from partially masked inputs, the ViT encoder gathers contextual information to predict the missing regions. This capability to aggregate context is especially important in medical imaging, where anatomical structures are functionally and mechanically linked to surrounding regions. However, current methods do not consider variations in the number of input images, which is typically the case in real-world Magnetic Resonance (MR) studies. To address this limitation, we propose a 3D Adaptive Masked Autoencoders (AMAE) architecture that accommodates a variable number of 3D input contrasts per subject. A magnetic resonance imaging (MRI) dataset of 45,364 subjects was used for pretraining and a subset of 1648 training, 193 validation and 215 test subjects were used for finetuning. The performance demonstrates that self pre-training of this adaptive masked autoencoders can enhance the infarct segmentation performance by 2.8\%-3.7\% for ViT-based segmentation models.

\end{abstract}
\section{Introduction}
\label{sec:intro}

In recent years, significant advancements have been made in applying deep learning methodologies to medical imaging tasks such as classification \cite{ismael2020medical,rahmat2018chest}, segmentation \cite{roth2018deep,malhotra2022retracted}, and object detection \cite{jaeger2020retina,li2019clu}. With the introduction of the Vision Transformer \cite{dosovitskiy2020image} and its success in computer vision tasks, researchers are now exploring different methodologies and applications in medical image analysis by leveraging transformers \cite{manzari2023medvit, das2024co, hatamizadeh2022unetr}. 

In UNETR \cite{hatamizadeh2022unetr}, the authors used a ViT encoder and a UNet \cite{ronneberger2015u} shaped decoder for medical image segmentation. Similarly, with the introduction of hierarchical Swin transformers \cite{liu2021swin}, researchers propsoed SwinUNETR \cite{hatamizadeh2021swin} for segmentation tasks. Zhou, Lei, et al \cite{zhou2023self} applied ViT-based masked autoencoders for self pre-training on medical image classification and segmentation tasks. A self-supervised pre-training approach can be used to enhance the downstream performance of ViT based architectures.

However, medical imaging presents unique challenges due to the complexity and variability of the data. Real world medical images often come in the form of multiple 3D scans, with varying number of contrasts for each acquisition, making it difficult to develop models that can handle this heterogeneity. Current models typically require the number of input images to be fixed. These approaches fall short of meeting the practical needs of medical diagnosis, where the number and types of input images can vary significantly from  acquisition to acquisition depending on the diagnostic task at hand and the protocols of different clinical sites.

To address these challenges, this work introduces a new flexible framework for medical imaging that enables the following:
\begin{itemize}
    \item The ability to handle a variable number of input images during both training and testing, improving the model's flexibility and adaptability to real world clinical data with varying contrasts.
    \item Self pre-training with variable number of 3D input contrasts and maximize data availability.
    
\end{itemize}

This framework aims to bridge the gap between current models and the complex demands of real-world medical diagnostics.

\section{Methodology}

\begin{figure*}[htb!]
    

  \centering
  \centerline{\includegraphics[width=12.5cm]{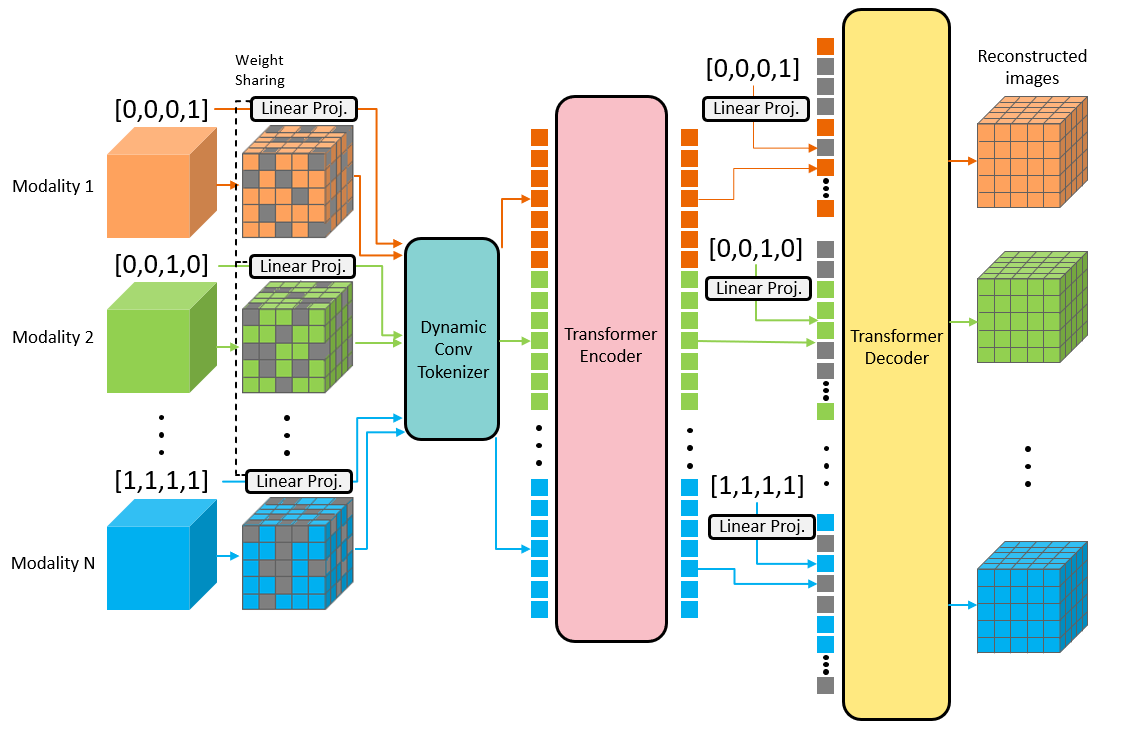}}
\caption{Adaptive MAE self-supervised pretraining architecture which can handle N number of modalities.}
\label{architecture}
\end{figure*}

\subsection{Data and Pre-processing}


 For pretraining we used 45,374 MR studies. For each case, a variable number of MRI contrasts acquired were used, including a set of seven contrasts: Apparent Diffusion Coefficient (ADC), Trace-weighted , T2-weighted, Gradient Echo (GRE), Susceptibility-Weighted , T1-weighted, and Fluid-Attenuated Inversion Recovery (FLAIR), or a subset of these when available. The images were acquired on scanners from Siemens Healthineers AG and, GE Medical Systems. For finetuning, We used a subset with 1648 training, 193 validation and 215 test subjects where acute/subacute brain infarct regions were manually segmented.



The manual segmentation of acute and subacute infarct lesions was performed on Axial Trace-weighted (AxTrace) contrast image series by a radiologist (T.J.R., 10 years of experience). The radiologist used the medical image segmentation software ITK-SNAP version 3.8.0. The AxTrace image series and corresponding ADC image map were loaded into the software and reviewed by the radiologist. Areas, within the brain parenchyma, of hyperintensity in the TraceW image series with hypo or iso-intensity in the ADC map were considered positive for recent (acute to subacute) infarct by the radiologist and delineated as such in an image mask using the software tool. The inclusion criteria were studies where at least AxTrace and AxADC contrasts are present and where the radiologist identified at least one infarct lesion by aforementioned method. Additionally, Axial T2-weighted contrast is used if it is available.


\subsection{Adaptive Masked Autoencoders}

The proposed adaptive framework employs a masked autoencoder approach for self-supervised learning (SSL) as shown in Figure \ref{architecture}. The process begins with random masking of a percentage of 3D patches in each modality’s original 3D volume. Inspired by the Dynamic Filter Network used in ModDrop++ \cite{liu2022moddrop++}, we propose a 3D Dynamic Convolution Tokenizer (DCT), which is then used to encode unmasked 3D patches into tokens. The DCT is designed to dynamically adapt to different input modalities.

Each modality is assigned a unique modality vector \( \mathbf{m}_i \) of length \( l \), where \( i \) represents the $i$-th modality:

\[
\mathbf{m}_i \in \mathbb{R}^l
\]

The modality vector \( \mathbf{m}_i \) is passed through a learnable linear projector and then split to generate a weight vector \( \mathbf{w}_{conv} \) and a bias vector \( \mathbf{b}_{conv} \). The weight vector \( \mathbf{w}_{conv} \) and bias vector \( \mathbf{b}_{conv} \) are used to update the weights and biases of the dynamic convolutional layer. Let \( \mathbf{W} \) and \( \mathbf{B} \) denote the original weights and biases of the convolutional layer. The updated weights \( \mathbf{W}_{updated} \) and biases \( \mathbf{B}_{updated} \) are:

\[
\mathbf{W}_{updated} = \mathbf{W} \cdot \mathbf{w}_{conv}
\]

\[
\mathbf{B}_{updated} = \mathbf{B} \cdot \mathbf{b}_{conv}
\]

This dynamic convolutional layer allows the tokenizer to extract modality-specific features from each 3D image. It also creates non overlapping tokens for the transformer by using patch size as kernel and stride size of the convolution and output dimension of the convolution as embedding dimension of the transformer. Let \( \mathbf{{X}_i} \in \mathbb{R}^{H \times W \times D} \) represent one 3D input image (or image tensor), where \( H, W, D \) are the height, width, and depth of the 3D image, respectively. The convolution operation with the updated weights and biases is:

\[
\mathbf{{Y}_i} = \text{Conv}(\mathbf{X}, \mathbf{W}_{updated}, \mathbf{B}_{updated})
\]

where \( \mathbf{{Y}_i} \in \mathbb{R}^{\left(\frac{H \times W \times D}{\text{Patchsize}^3}\right) \times \text{EmbeddingDim}} \) is the output tokens produced by the dynamic convolutional layer for one input image.




The unmasked tokens from all modalities are then concatenated into a long sequence and then fed into a transformer encoder. Each token receives a positional embedding (either sinusoidal or learnable) based on its 3D patch’s location in the 3D volume as transformer is position-agnostic. Two tokens from different modalities but the same patch location will have the same positional embedding. These embeddings help preserve spatial relationships of the tokens. 


During pretrain, the transformer encoder only encodes the unmasked tokens. Then, a placeholder token is inserted to the output of the transformer encoder at each maksed token’s position. Both masked and unmasked tokens are then added with the positional and modality embedding again and fed into a Transformer decoder. A linear projector then maps the decoded tokens back to the size of the 3D patches to reconstruct the original 3D volume of all given input contrasts

\subsection{Downstream Architecture}
For finetuning for the task of infarct segmentation we have two approaches. For the first approach, referred to as Adaptive UNETR, we keep the DCT and Transformer encoder as it is and remove the Transformer decoder and then use a UNETR based segmentation decoder head. UNETR is a widely used UNet shaped transformer model for 3D medical image segmentation. However, as we have variable number of inputs for each subject our number of tokens also vary. We use an adaptive max pooling layer to extract the necessary information across all the modalities at each level in the UNETR decoder head to transfer variable number of tokens to tensors with fixed sizes to generate the segmentation masks. 

Let \( \mathbf{X} \) denote the output of the transformer encoder:

\[
\mathbf{X} \in \mathbb{R}^{B \times N \times \left(\frac{H \times W \times D}{\text{Patchsize}^3}\right) \times \text{EmbeddingDim}}
\]

where \( B \) is the batch size, \( N \) is the number 3D input images, \(\left(\frac{H \times W \times D}{\text{Patchsize}^3}\right)\) is the number the patches/tokens, \(\text{EmbeddingDim}\) is the transformer dimension.

We use an adaptive max pooling layer to extract the necessary information from all tokens, resulting in:

\[
\mathbf{X}_{new} \in \mathbb{R}^{B \times \left(\frac{H \times W \times D}{\text{Patchsize}^3}\right) \times \text{EmbeddingDim}}
\]

In another approach, we used the original UNETR model as it is. All the images from different modalities are concatenated along the channel dimension. Zero-filled tensor is used if any input modality is missing. We transfer the weights of the pretrained ViT encoder from our pretraining to the UNETR encoder.

    


\subsection{Training and Evaluation}

The initial learning rate for self pre-training is 1e-5 and weight decay 0.05. We use a L2 loss function and weighted Adam optimizer. For pre-training we run our variable masked autoencoder model till 500 epochs with a masking ratio of 70\% and batch size of 4. 

For finetuning for the task of infarct segmentation we run the finetune architecture till 200 epochs and saved the best validation result model. Later we used test dataset to compare the performance of different models. We compared the performance of our model with UNet \cite{ronneberger2015u} and UNETR \cite{hatamizadeh2022unetr}. The learning rate for finetuning was 0.0001 with learning rate scheduler and weight decay of 0.05. We used Diceloss and weighted ADAM optimizer for training.  Additionally, we performed a test where we evaluated model performance on the test data by omitting the optional contrast AxT2, to observe how the models behave when a contrast/modality is missing.

Dice similarity coefficient was used for quantitative evaluations on the test data. Two-sided pairwise Wilcoxon signed rank test was used to compare the Dice scores of two models with and without pre-training. The experiments were implemented using PyTorch(v1.12.1) and the Monai14(v1.1.0) framework.

\section{Results}
\begin{figure}[htb!]
    

  \centering
  \centerline{\includegraphics[width=1.0\linewidth]{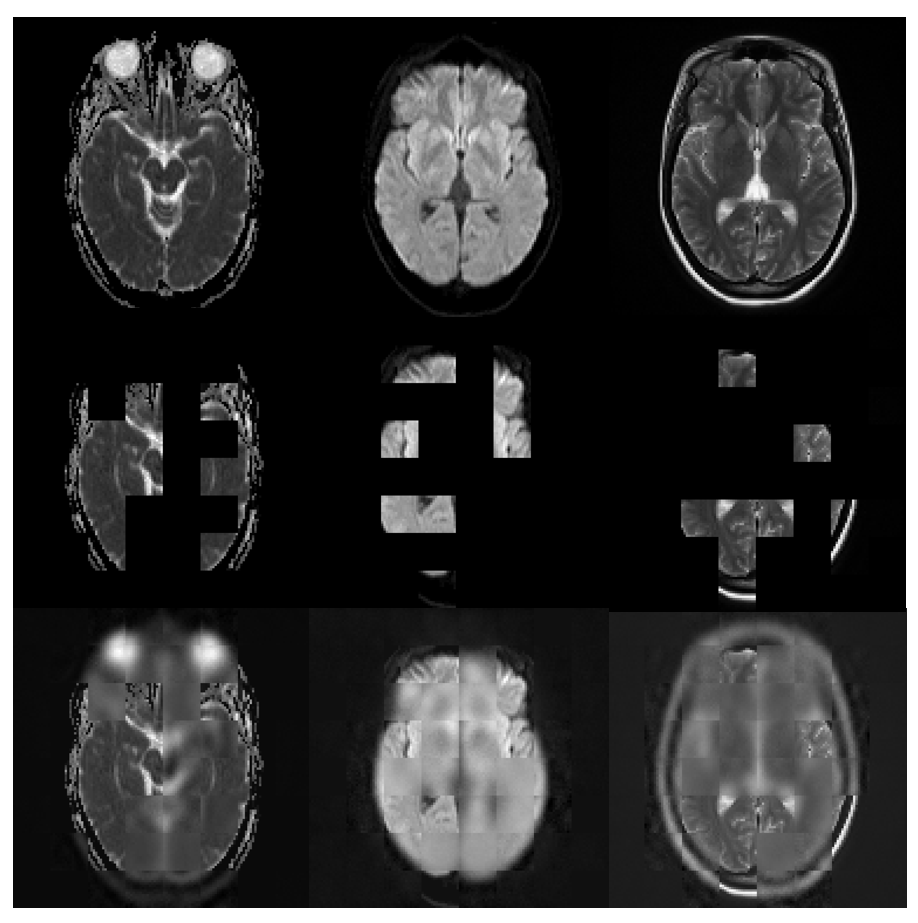}}
\caption{Reconstruction of  pre-training. First row: original image. Second row: masked image where masked patches are colored as black. Third row: reconstructed images. (Axial slices of ADC, Trace and T2 are shown from left to right)}
\label{pretrain_rec}
\end{figure}

\begin{table}[h!]
\centering
\begin{tabular}{|c|c|}
\hline
\textbf{Model} &   \textbf{Mean Dice Score} \\ \hline
UNET  & 0.370   \\ \hline
UNETR  & 0.576\\ \hline

Adaptive UNETR & 0.561 \\ \hline

AMAE pretrain + Adaptive UNETR  & 0.598  \\ \hline

AMAE pretrain + UNETR & 0.604  \\ \hline

\end{tabular}
\caption{Performance comparison of different models using mean Dice score on test data.}
\label{table:performance_metrics}
\end{table}

The pretraining reconstruction of ADC, Trace-weighted and T2-weighted images are shown in Figure \ref{pretrain_rec}. The three rows consist of a single slice of the original input image of AxADC, AxTrace and AxT2 contrasts, their corresponding masked images and the reconstructed images, respectively. The outcomes show that MAE can recover the deleted data from the random context. The recovered visible patches appear blurry because the L2 loss is applied exclusively to the masked patches. However, it is a known phenomenon with self-pretraining of 3D medical images \cite{zhou2023self} and it is important to note that MAE does not aim to produce high-quality reconstructions but to extract useful features to aid downstream tasks.

\begin{table}[ht]
\centering
\begin{tabular}{|c|c|}
\hline
\textbf{Model}  & \textbf{Mean Dice Score} \\ \hline

Unet  & 0.343 \\ \hline
UNETR & 0.516 \\ \hline

Adaptive UNETR & 0.496 \\ \hline

AMAE pretrain + Adaptive UNETR  & 0.546\\ \hline

AMAE pretrain + UNETR & 0.575  \\ \hline

\end{tabular}
\caption{Performance of different models on 215 subjects using mean Dice score, where AxT2 absent for all subjects.}
\label{table:test_dice_axt2}
\end{table}



Effectiveness of the proposed AMAE pretrain with adaptive UNETR and the original UNETR are presented in Table \ref{table:performance_metrics}. The adaptive UNETR with pretrained weights performed 3.7\% better than the adaptive UNETR without the AMAE pretrained weights. Also, in case of the original UNETR the performance improved by 2.8\% after using weights from AMAE pretraining.

The results in Table \ref{table:test_dice_axt2} demonstrate the performance of different models on test subjects using the mean Dice score when the optional contrast AxT2 is absent. AMAE pretrained weights with the original UNETR decoder achieved mean Dice score of 0.575 with only 2.9\% lower than its performance with all contrasts.


In Table \ref{table:stats}, statistical analysis on Dice score of the two finetune approaches (Adaptive UNETR and UNETR) with pretraining and without pretraining are shown. In both cases, P-value is less than 0.05 which indicate significant difference in segmentation results while applying the pretrained weights compared to the corresponding model without pretrained weights.




\begin{table}[htb!]
\centering
\begin{tabular}{|c|c|}
\hline
\textbf{Model Comparison} & \textbf{P-Value} \\ \hline
\makecell{AMAE pretrain + Adaptive UNETR \\ vs. Adaptive UNETR} & $<0.05$ \\ \hline
\makecell{AMAE pretrain + UNETR \\ vs. UNETR} & $<0.05$ \\ \hline
\makecell{AMAE pretrain + Adaptive UNETR \\ vs. AMAE pretrain + UNETR} & $0.47$ \\ \hline
\makecell{Adaptive UNETR \\ vs. UNETR} & $0.32$ \\ \hline
\end{tabular}
\caption{Statistical Comparison of finetune models with and without AMAE pretrained weights.}
\label{table:stats}
\end{table}

\section{Discussion}

In this paper, we proposed a new pre-training and finetune framework which is able to accept variable number of input modalities. The pretraining weights enhanced the performance of both our finetune architecture as well as the original UNETR architecture. 


Another notable observation from the results is that generating patches from all input images together leads to slightly better performance in the fine-tuning task. The proposed AMAE pretrain weights can still be applied in this setting to further enhance performance, as demonstrated in Table \ref{table:performance_metrics} and Table \ref{table:test_dice_axt2}. The performance difference between UNETR and our Adaptive UNETR approach is not statistically significant, both with pretraining (P-value = 0.47) and without pretraining (P-value = 0.32).

Several limitations should be noted. This work primarily aims to enhance the flexibility of the architecture while achieving results comparable to other state-of-the-art models. We exclusively used the base ViT for all Transformer experiments and pretraining, though utilizing larger or more extensive ViT versions could potentially boost performance. Additionally, our pretraining was conducted with a fixed masking ratio of 0.7. In the future, we plan to explore different ViT dimensions and varying masking ratios to extend these experiments.

\section{Conclusion}

In conclusion, self pre-training of the proposed adaptive masked autoencoders enhances the performance of Vision Transformer models while handling varying 3D input contrasts. This method opens up a new way to handle heterogeneous real world clinical datasets and can be used for creation of 3D foundation model for medical imaging.

\section{Compliance with ethical standards}
\label{sec:ethics}

This retrospective study is compliant with the health insurance portability and accountability act (HIPAA). The dataset was collected and anonymized from five different centers; each hospital’s institutional review board approved this study for human research with waiver of informed consent. All methods are performed in accordance with the relevant guidelines and regulations.   


\section{Acknowledgments}
\label{sec:acknowledgments}

This research project was funded by Siemens Healthineers. We acknowledge the usage of MRI images from the Mount Sinai Hospital. The concepts and information presented in this paper/presentation are based on research results that are not commercially available. Future commercial availability cannot be guaranteed.



\bibliographystyle{IEEEbib}
\bibliography{strings,refs}

\end{document}